\documentclass[sigconf,screen]{acmart}

\AtBeginDocument{%
  }

\setcopyright{acmlicensed}
\copyrightyear{2026}
\acmYear{2026}
\acmDOI{XXXXXXX.XXXXXXX} 
\acmConference[FSE 2026]{ACM International Conference on the Foundations of Software Engineering}{July 05--09, 2026}{Montreal, Canada}

\acmBooktitle{Companion Proceedings of the 34th ACM Symposium on the Foundations of Software Engineering (FSE '26), June 5--9, 2026, Montreal, Canada}
\acmISBN{c978-1-4503-XXXX-X/2018/06}

\usepackage[table]{xcolor} 
\usepackage{tikz}
\usepackage{multirow}
\usepackage{diagbox}
\usetikzlibrary{positioning, fit, shapes}

\usepackage{tablefootnote}
\usepackage{listings}
\usepackage[edges]{forest}
\usetikzlibrary{positioning}
\usepackage{tcolorbox}

\usepackage{tcolorbox}
\tcbuselibrary{skins, breakable, theorems}

\definecolor{purplish}{HTML}{D8DFE3}
\definecolor{purplishlight}{HTML}{EBEFF3}
\definecolor{purplishdark}{HTML}{009e73}

\newtcolorbox[auto counter,number within=section]{rqbox}[2]{
    nameref=#1,
    title=\small{#1}, 
    enhanced,
    attach boxed title to top left={yshift=-6pt, xshift=8pt},
    boxed title style={size=small,boxsep=1pt},
    colframe=purplishdark,colback=white,colbacktitle=purplishdark,
    boxsep=2pt,left=2pt,right=2pt,top=6pt,bottom=2pt,middle=2pt
}

\newcommand{\rqtextone}{How do software engineering students describe their experiences with LLM cheating, and what contextual, motivational, and educational factors influence these behaviors?}

\begin{document}

\title{LLM Use, Cheating, and Academic Integrity in Software Engineering Education}

\author{Ronnie de Souza Santos}
\affiliation{
  \institution{University of Calgary}
  \city{Calgary}
  \country{Canada}
}
\email{ronnie.desouzasantos@ucalgary.ca}

\author{Italo Santos}
\affiliation{
  \institution{University of Hawaii at Manoa}
  \city{Honolulu}
  \country{USA}
}
\email{isantos3@hawaii.edu}

\author{Mariana Bento}
\affiliation{
  \institution{University of Calgary}
  \city{Calgary}
  \country{Canada}
}
\email{mariana.pinheirobent@ucalgary.ca}

\author{Giuseppe Destefanis}
\affiliation{
  \institution{University College London}
  \city{London}
  \country{United Kingdom}
}
\email{g.destefanis@ucl.ac.uk}

\author{Cleyton Magalhães}
\affiliation{
  \institution{Universidade Federal Rural de Pernambuco (UFRPE)}
  \city{Recife}
  \country{Brazil}
}
\email{cleyton.vanut@ufrpe.br}

\author{Mairieli Wessel}
\affiliation{
  \institution{Radboud University}
  \city{Nijmegen}
  \country{Netherlands}
}
\email{mairieli.wessel@ru.nl}

\begin{abstract}
\textit{Background:} Cheating in university education is commonly described as context-dependent and influenced by assessment design, institutional norms, and student interpretation. In software engineering education, programming-oriented coursework has historically involved ambiguity around collaboration, reuse, and external assistance. Recently, large language models (LLMs) have introduced additional mediation in the production of code and related artifacts. \textit{Aims:} This study investigates how software engineering students describe experiences of using LLMs in ways they perceived as inappropriate, disallowed, or misaligned with course expectations. \textit{Method:} A cross-sectional survey was conducted with 116 undergraduate software engineering students from multiple countries, combining quantitative summaries with qualitative data. \textit{Results:} Reported LLM cheating practices occurred primarily in programming assignments, routine coursework, and documentation tasks, often in contexts of time pressure and unclear guidance. Use during quizzes and exams was less frequent and more consistently identified as a violation. Students reported awareness of academic and professional consequences regarding LLM cheating, while formal sanctions were perceived as limited. \textit{Conclusions:} Our study indicates that reported LLM misuse in software engineering is associated with assessment and instructional conditions, suggesting a need for clearer alignment between assessment design, learning objectives, and expectations for LLM use.
\end{abstract}



\received{22 January 2026}

\maketitle

\section{Introduction}
\label{intro}

Cheating in university settings is commonly defined as behavior that violates explicit assessment rules or established academic norms, rather than a single, clearly bounded act~\cite{whitley1998factors, mccabe2001cheating, sheard2002cheating}. Prior research characterizes cheating as a spectrum of practices, ranging from unauthorized collaboration and plagiarism to exam misconduct and impersonation~\cite{franklyn1995undergraduate, magnus2002tolerance}. What students consider cheating is context-dependent and influenced by assessment design, instructional guidance, and peer norms~\cite{hutton2006understanding, molnar2012does}. Consequently, lower severity practices are sometimes viewed as acceptable, while more explicit violations are recognized as serious but reported less frequently~\cite{sheard2002cheating, jones2011academic}. Previous studies indicate that academic misconduct is widespread rather than confined to a small subset of students, with many reporting at least one form of cheating during their studies~\cite{whitley1998factors, mccabe2001cheating}. Prior behavior, perceptions of peer conduct, and expectations of detection are repeatedly associated with continued misconduct~\cite{whitley1998factors, magnus2002tolerance}, while demographic factors such as age and academic progression show mixed but sometimes inverse associations with cheating frequency~\cite{franklyn1995undergraduate, whitley1998factors}.

In software engineering and computing programs, cheating reflects discipline-specific characteristics of programming work and software artifacts. Research reports relatively high tolerance for practices such as collaboration on individual assignments and reuse of existing code, particularly among early-stage students~\cite{sheard2002cheating}. Comparative studies suggest that self-reported cheating tends to decrease with academic maturity and professional orientation, although it does not disappear entirely~\cite{sheard2003investigating}. In programming courses, common forms of misconduct include copying or adapting peers’ code, reusing solutions from prior offerings, and relying on external assistance beyond what is permitted~\cite{salhofer2017analysing, alzahrani2022detecting}. These practices are often interpreted as emerging from the cumulative and material nature of software development, where reuse is technically straightforward but pedagogically constrained.

The introduction of generative AI and large language models (LLMs) has changed cheating practices at the university, primarily by modifying how assistance is accessed rather than by introducing new motivations~\cite{cotton2024chatting, adnan2025cheating}. Students continue to balance effort, risk, and norms when deciding whether to engage in misconduct, but LLMs substantially reduce the effort required to generate plausible outputs~\cite{firth2024cheating, lee2024cheating}. This shift increases ambiguity around the boundary between acceptable support and cheating, especially when assessment policies or learning objectives are unclear~\cite{mah2024beyond, ortiz2025chat}. Detection-focused responses face limitations, as AI-generated artifacts complicate judgments of originality, authorship, and intent~\cite{spennemann2024chatgpt, kundu2024keystroke}. In software engineering education specifically, cheating with LLMs is more visible in short, well-specified programming assignments focused on functional correctness, for which widely available models can often produce correct or near correct solutions~\cite{rytilahti2024easy, salim2024impeding}. Prior work indicates that changes in assignment structure can reduce the effectiveness of such practice~\cite{salim2024impeding}, while conceptual analyses frame the phenomenon as a systemic issue linked to misalignment between assessment practices, learning goals, and professional expectations~\cite{randall2023ai, marlowe2026systemic}.

Building on this body of work, the goal of this study is to investigate software engineering students’ lived experiences with LLM use that they perceive as inappropriate, disallowed, or misaligned with course expectations. Rather than focusing on prevalence, the study examines how students describe the circumstances that led to such use, how they interpret or justify their decisions, and how they understand the associated risks and consequences of cheating with LLMs. This focus responds to calls to move beyond detection-centered approaches toward closer analysis of student reasoning and decision making in AI-mediated learning environments~\cite{mah2024beyond, lee2024cheating}. In this sense, this study addresses the following research question: 

 \newcommand{\rqone}[2][]{
     \begin{rqbox}{\textbf{Research Question}}{#2}
         \rqtextone
         #1
     \end{rqbox}
 }

 \rqone{}


The novelty of this study lies in its emphasis on students’ own accounts of LLM-assisted cheating, addressing a gap in prior literature that has largely documented prevalence, task-level vulnerabilities, or technical and policy responses from instructor or institutional perspectives~\cite{sheard2002cheating, salim2024impeding, cotton2024chatting}. From this introduction, the remainder of this paper is organized as follows. Section~\ref{sec:background} reviews relevant literature on academic cheating. Section~\ref{sec:method} describes the study design and methodological procedures. Section~\ref{sec:findings} presents the results, which are discussed in Section~\ref{sec:discussion}. Section~\ref{sec:conclusion} summarizes the contributions and highlights implications for research and educational practice.

\section{Background}\label{sec:background}

This section reviews prior literature on cheating and academic misconduct in higher education, focusing on how practices, interpretations, and responses have evolved over time.

\subsection{Evolving Cheating Practices in the University}

Early research from Education characterizes cheating as a persistent and routine feature of university education, occurring across disciplines and assessment types rather than as an exceptional behavior. Studies document a wide range of cheating practices, including copying during exams, unauthorized collaboration, plagiarism, and submission of others’ work, with students commonly distinguishing between behaviors viewed as minor versus serious violations \cite{franklyn1995undergraduate, whitley1998factors, mccabe2001cheating}. The literature indicates that cheating was embedded in everyday academic practices and shaped by peer norms, institutional context, and assessment design. Students’ interpretations of what constituted cheating vary substantially depending on the academic setting, even when formal rules were in place, and attitudes toward specific practices differed across contexts, influenced by perceived fairness, instructor practices, peer behavior, and expectations of punishments \cite{magnus2002tolerance, hosny2014attitude, bernardi2004examining, hutton2006understanding}. Several studies also suggest that contextual and social factors were more consistently associated with cheating than individual characteristics, and that institutional responses such as honor codes were effective primarily when supported by shared norms rather than by formal sanctions alone \cite{whitley1998factors, mccabe2001cheating}.

As digital technologies and online resources became more available, researchers began to investigate how these tools mediated existing cheating practices \cite{jones2011academic, hsiao2015impact}. Studies on internet plagiarism and technology-supported learning environments report increased opportunities for copying, paraphrasing, and reuse of online materials, often accompanied by student uncertainty regarding citation and attribution norms \cite{jones2011academic, hsiao2015impact}. This body of work does not suggest a fundamental change in students’ motivations, but rather an expansion of the means through which established practices could be enacted \cite{whitley1998factors, hutton2006understanding}. Institutional responses increasingly emphasized detection-oriented mechanisms such as plagiarism detection and automated monitoring, although concerns were raised about uneven enforcement and students’ capacity to adapt to these systems \cite{hutton2006understanding}. Systematic reviews have documented a strong institutional focus on detection and control as primary responses to technology-mediated misconduct \cite{noorbehbahani2022systematic}.

More recent literature on generative AI and large language models situates current concerns within this longer trajectory of technological mediation and cheating \cite{cotton2024chatting, adnan2025cheating, mah2024beyond}. Empirical and conceptual studies suggest that LLMs affect cheating practices by reducing the effort required to produce acceptable academic outputs and by increasing ambiguity around permissible assistance \cite{cotton2024chatting, adnan2025cheating}. Other studies indicate that AI-generated content can satisfy formal assessment criteria without demonstrating understanding, raising questions about originality, authorship, and independent work \cite{firth2024cheating}. Students’ accounts indicate continuity with earlier models of cheating, as decisions remain shaped by perceived risk, norms, and effort rather than by the technology alone \cite{lee2024cheating, ortiz2025chat}. Across this literature, concerns persist that the consequences of cheating extend beyond immediate academic penalties to include reduced learning, weakened trust in assessment systems, and potential implications for professional preparation and ethical reasoning \cite{whitley1998factors, mccabe2001cheating, cotton2024chatting}.

\subsection{Academic Integrity and Cheating in Software Engineering Education}

Software engineering education is commonly organized as a cumulative curriculum that progresses from foundational computing skills toward applied engineering practice. Academic programs typically begin with programming, data structures, and algorithms, and advance to activities such as requirements analysis, design, testing, maintenance, and project management, often culminating in team-based projects and capstone experiences \cite{randall2023ai, marlowe2026systemic}. In software engineering, assessments often rely on the production of software artifacts such as code, documentation, and integrated systems. Prior educational research characterizes this artifact-centered structure as creating pedagogical tensions, since reuse, modification, and incremental development are standard professional practices but are often restricted in educational settings to support learning and avoid cheating \cite{sheard2002cheating, randall2023ai, marlowe2026systemic}.

Before the introduction of large language models, cheating in software engineering education was studied as a discipline-specific phenomenon shaped by programming-oriented assessment and artifact-based coursework \cite{sheard2002cheating, sheard2003investigating, salhofer2017analysing}. Previous studies emphasize student uncertainty around collaboration, code reuse, and individual responsibility, particularly in early stages of study, reflecting ambiguity in rule interpretation rather than agreement about misconduct \cite{sheard2002cheating}. Some authors suggest that reported cheating decreases with academic progression but remains present, and that programming-related misconduct is commonly identified through patterns of reuse and code similarity \cite{sheard2003investigating, salhofer2017analysing, alzahrani2022detecting}. More recent work associates the availability of large language models with an extension of these established patterns through automated code generation and modification, raising concerns about reduced engagement in core engineering activities, interpretation of artifacts as evidence of learning, and broader misalignment between assessment practices and expectations of professional competence \cite{salhofer2017analysing, rios2023authorship, rytilahti2024easy, salim2024impeding, randall2023ai, marlowe2026systemic}.

\section{Method} \label{sec:method}

This study focuses on software engineering students’ experiences with the use of LLMs in ways they perceived as inappropriate, disallowed, or misaligned with course expectations. Building on prior empirical work on academic cheating and, more recently, LLM-assisted cheating in software engineering and higher education that relies on self-reported data to investigate student practices and decision-making~\cite{sheard2002cheating, sheard2003investigating, hosny2014attitude, lee2024cheating, cotton2024chatting, adnan2025cheating, ortiz2025chat}, we adopted a cross-sectional survey design~\cite{pfleeger2001principles, easterbrook2008selecting, ralph2020empirical}. The survey combined closed-ended and open-ended questions to capture patterns of LLM use as well as students’ accounts of how they interpreted rules, justified their decisions, and understood potential consequences. The subsections below describe the survey instrument, participant recruitment and sampling, data collection procedures, and data analysis approach.

\subsection{Survey Design}

Following the research guidelines~\cite{pfleeger2001principles, linaker2015guidelines}, we developed an anonymous survey to capture software engineering students’ self-reported experiences with the use of LLMs in ways they perceived as conflicting with course expectations, assessment rules, or academic integrity policies. The survey was implemented using Qualtrics\footnote{www.qualtrics.com} and structured into five sections, followed by a demographic and contextual section. The instrument combined open-ended questions with multiple-choice items to support both descriptive characterization and in-depth accounts of student reasoning.

The survey was explicitly designed to capture students’ lived or observed experiences with LLM use that conflicted with course expectations. Concrete instances of such use, including the types of coursework involved and the forms of assistance employed, were documented alongside motivations, constraints, justifications, and emotional responses. Perceptions of where these practices occurred within software engineering programs and which curricular areas were most affected were also documented, together with students’ views on academic, personal, and professional consequences and potential deterrents. Closed-ended questions supported the identification of recurring patterns related to assessment context, perceived risk, workload, peer norms, and institutional guidance, while open-ended prompts elicited detailed interpretations and reasoning. Background information was collected to contextualize responses while preserving anonymity, and no personally identifiable information was gathered. Table~\ref{tab:surveyquestions} presents the full set of survey questions.

\begin{table}[t]
\caption{Survey Questionnaire}
\label{tab:surveyquestions}
\centering
\scriptsize
\begin{tabular}{p{8cm}}
\hline
\textbf{EXPERIENCES WITH INAPPROPRIATE LLM USE} \\

1. Describe a situation in your software engineering program where you used an LLM in a way that you believe was not aligned with the course rules or expectations.  \\ \\

2. In what types of coursework have you used an LLM in ways that were likely not permitted? (Select all that apply)
( ) Regular classwork or weekly exercises / ( ) Programming assignments or labs / ( ) Group or capstone projects / ( ) Essays or written reflections / ( ) Software design or documentation tasks / ( ) Quizzes or short online tests / ( ) Major exams or finals / ( ) Other. \\ \\

3. Which of the following LLM uses have you personally engaged in, even if they were not allowed by the course? (Select all that apply)
( ) Submitting LLM-generated code or solutions as my own work / ( ) Copying LLM-generated text into reports or essays without acknowledgment / ( ) Using LLMs during quizzes, tests, or exams / ( ) Having LLMs perform debugging or test-case generation for assessed work / ( ) Asking LLMs to produce substantial parts of group project deliverables / ( ) Using multiple LLMs or rewriting strategies to obscure AI involvement / ( ) Other use that conflicted with course expectations. \\ \\

4. Which factors made this type of LLM use easier or more tempting? (Select all that apply)
( ) Assignments completed remotely or online / ( ) Unclear or missing rules about AI use / ( ) Limited oversight or checking by instructors / ( ) Heavy workload or overlapping deadlines / ( ) Peer norms encouraging similar behavior / ( ) Belief that AI use would not be detected / ( ) Other. \\ \hline

\textbf{MOTIVATIONS AND DECISION FACTORS} \\

5. What were your main reasons for using an LLM in this way? \\ \\

6. What factors contributed to your decision to use an LLM in ways that were not permitted? (Select all that apply) ( ) Pressure to achieve high grades / ( ) Too many assignments or overlapping deadlines / ( ) Tasks that felt too difficult or poorly specified / ( ) Low confidence in my own skills / ( ) Perception that many peers were doing the same / ( ) Low perceived risk of consequences / ( ) Initial experimentation that escalated / ( ) Desire to save time or reduce stress. \\ \\

7. How did you interpret or justify this decision at the time? ( ) I felt it was necessary to manage the workload / ( ) I believed the rules were unclear or outdated / ( ) I thought the impact was minor / ( ) I avoided thinking about possible consequences / ( ) I knew it violated expectations but accepted the risk. \\ \\

8. Afterward, which emotion best describes how you felt? ( ) Relief / ( ) Guilt / ( ) Indifference / ( ) Satisfaction / ( ) Anxiety about consequences. \\
\hline

\textbf{AFFECTED AREAS OF THE PROGRAM} \\

9. In your experience, which aspects of software engineering education are most affected by inappropriate or unclear LLM use? (Select all that apply) ( ) Programming and implementation / ( ) Software testing or debugging / ( ) System design or architecture / ( ) Project management and teamwork / ( ) Technical documentation or reports / ( ) Individual skill assessments. \\ \\

10. Which course contexts made it easiest to rely on LLMs in ways that conflicted with expectations? ( ) Introductory programming courses / ( ) Advanced software engineering or design courses / ( ) Testing and quality assurance courses / ( ) Capstone or group projects / ( ) Theory or writing-focused courses such as ethics or requirements. \\
\hline

\textbf{PERCEIVED CONSEQUENCES AND DETERRENTS} \\

11. What personal, academic, or professional consequences do you think may result from relying on LLMs in ways that bypass learning or assessment goals? \\ \\

12. If such LLM use were formally identified, how serious do you believe the consequences would be? \\

13. Even if no formal consequences occur, what do you see as the main negative effects?\\ \\

14. What would most discourage you from using LLMs inappropriately in the future? (Select all that apply) ( ) Clearer and more consistent policies / ( ) Assessments involving oral explanation or code walkthroughs / ( ) Better alignment between workload and expectations / ( ) Assignments that explicitly integrate acceptable AI use / ( ) Stronger penalties for misuse. \\
\hline

\textbf{DEMOGRAPHIC} \\

15. Year in your program? \\ \\

16. Have you received any formal training or guidance on responsible LLM use in your program? ( ) Yes / ( ) No / ( ) Not sure. \\ \\

17. In the courses where this occurred, were expectations about LLM use communicated? ( ) Yes, clearly and explicitly / ( ) Yes, but only in general terms / ( ) No guidance was provided / ( ) I do not remember. \\ \\

18. Were potential consequences of inappropriate LLM use discussed? ( ) Yes, clearly / ( ) Yes, briefly / ( ) No / ( ) Not sure. \\ \\

19. Does your institution have a formal policy on AI or LLM use in coursework? ( ) Yes / ( ) Not that I know of / ( ) Not sure. \\ \\

20. What best describes your gender identity? \\ \\

21. Country where you are currently studying? \\
\hline
\end{tabular}
\end{table}

\subsection{Pilot}
The survey instrument underwent a pilot phase after initial drafting and internal review. The pilot served both technical and content-related purposes. First, the survey was tested across different browsers and devices to verify functionality, navigation, and response recording. Second, a content pilot was conducted with two software engineering faculty members and three undergraduate students enrolled at different stages of a software engineering program. Pilot participants were asked to complete the survey and provide feedback on the clarity, interpretability, and relevance of the questions. Their input informed revisions to wording, ordering, and response options, with particular attention to ensuring that prompts were unambiguous and encouraged reflection on personal experiences rather than generalized opinions. Feedback also supported refinement of the balance between open-ended and closed-ended questions.

\subsection{Recruitment}
Participant recruitment followed established guidance for empirical software engineering research and relied on a combination of platform-based sampling and referral-based snowball sampling~\cite{ralph2020empirical, baltes2022sampling}. The primary recruitment channel was the Prolific platform, which supports structured prescreening and has been widely adopted in software survey-based studies~\cite{russo2022recruiting, reid2022software}. Prolific filters were configured to identify individuals who reported current enrollment in undergraduate software engineering or closely related programs and prior experience using LLMs in academic coursework. These platform-level criteria were complemented by eligibility questions embedded in the survey to verify consistency. To extend participation beyond the Prolific participant pool, we additionally employed snowball sampling~\cite{baltes2022sampling}. Participants who completed the survey were invited to share the study with peers enrolled in similar programs who met the eligibility criteria. This approach helped increase coverage. Data quality considerations were integrated throughout the recruitment process. Participants were required to confirm their field of study, level of enrollment, and experience with LLMs before proceeding. Data collection remained open for two weeks, resulting in a sample that included students from different academic years and institutional contexts.

\subsection{Filtering}
Following data collection, the dataset was subjected to a multi-stage filtering process to ensure validity and data quality, following recommendations for survey-based software engineering research~\cite{danilova2021you, alami2024you}. First, responses were screened against the eligibility criteria using both Prolific prescreening attributes and survey-level validation questions. Incomplete submissions and responses that failed these checks were removed. Second, attention and engagement checks were applied. Responses that failed attention checks or exhibited uniform answering patterns across closed-ended questions were excluded. Completion times were also observed, and responses with unusually short durations were removed, as these suggested insufficient engagement with the survey content. Finally, given the focus on students’ lived experiences and personal reasoning, responses were also reviewed for indicators of automated or AI-generated text. Submissions characterized by repetitive phrasing, generic statements lacking situational detail, or stylistic patterns inconsistent with reflective self-reporting were excluded. This step aimed to ensure that the dataset reflected participants’ own accounts rather than AI-generated content.

\subsection{Data Analysis}
The survey produced both qualitative and quantitative data. Open-ended responses were analyzed using an iterative thematic analysis approach, following established qualitative research guidelines in software engineering~\cite{cruzes2011recommended}. Analysis began with an inductive coding phase, in which responses were read in full to identify salient ideas, situations, and forms of reasoning articulated by participants. These initial codes captured recurring patterns related to contexts of LLM use, motivations, interpretations of course rules, and perceived consequences. In subsequent stages, conceptually related codes were grouped into higher-level themes aligned with the study’s research question, and then refined to reduce redundancy, clarify thematic boundaries, and ensure coherence across the codebook. All coding was conducted by at least two researchers. Each researcher independently coded a subset of responses, after which discrepancies were discussed in joint review sessions and resolved through consensus, supporting analytical consistency and reducing individual interpretive bias. Closed-ended items and demographic variables were analyzed using descriptive statistics~\cite{george2018descriptive}. These summaries characterized distributions for coursework type, perceived risk, deterrents, and participant background and were used to contextualize and support interpretation of the qualitative findings rather than to enable inferential analysis. 

\subsection{Ethics}
The study followed institutional guidelines for research involving human participants and received approval from the first author's university ethics board. Informed consent was obtained before participation, and respondents were informed about the study’s purpose, voluntary nature, data handling practices, and their right to withdraw. No personally identifiable information was collected, and responses were analyzed and reported in aggregate to preserve anonymity. These safeguards were particularly important given the study’s focus on experiences with potentially disallowed or sensitive academic practices, where protecting participant confidentiality was necessary to reduce risk, encourage honest reporting, and ensure ethical data collection.

\section{Results}
\label{sec:findings}

We first report a demographic summary of the participants to contextualize the dataset. We then present findings on how the use of LLMs in ways perceived as inappropriate or misaligned with course expectations occurs in practice, followed by an analysis of students’ reported motivations and reflections on these practices. Throughout the presentation of results, we include quotations from participants to illustrate and substantiate the findings. These quotations are reproduced verbatim, and as a result, some may read awkwardly or contain informal phrasing. This decision was made deliberately to preserve the original meaning of participants’ accounts and to support transparency and analytical validity. \\

\noindent \textbf{Participant Demographics.} 
The survey includes responses from 116 undergraduate students currently enrolled in higher education institutions across multiple countries. Participants were studying in a broad range of geographic contexts, with the largest representation from South Africa (19.8\% – 23 participants), the United Kingdom (10.3\% – 12 participants), the United States of America (9.5\% – 11 participants), Canada (8.6\% – 10 participants), and Germany (6.9\% – 8 participants). Additional participants were based in Italy (5.2\% – 6), Hungary and India (4.3\% – 5 each), France, Mexico, the Philippines, Poland, and Portugal (2.6\% – 3 each), and Brazil, Chile, Greece, and Latvia (1.7\% – 2 each). Smaller representations were observed from Australia, Austria, the Czech Republic, Egypt, Ireland, Kenya, the Netherlands, New Zealand, Sweden, and Vietnam (0.9\% – 1 each). In terms of year of study, participants were distributed across different stages of their undergraduate programs. First-year students accounted for 19.8\% (23 participants), second-year students for 16.4\% (19 participants), third-year students for 28.4\% (33 participants), and fourth- or fifth-year students for 35.3\% (41 participants). This distribution suggests a stronger representation of students in later stages of their undergraduate studies. Finally, regarding gender identity, 62.9\% (73 participants) identified as men, 36.2\% (42 participants) identified as women, and 0.9\% (1 participant) identified as non-binary. Gender was self-reported, and no additional categories were selected. \\

\noindent \textbf{Institutional Guidance on LLM Use.}
Participants reported varying levels of institutional and course-level guidance regarding the use of LLMs. When asked whether they had received formal training or guidance on responsible LLM use within their program, 54.3\% (63 participants) indicated that they had not received any such training, while 43.1\% (50 participants) reported that they had received some form of formal guidance. A small proportion of participants 2.6\% (3 participants) were unsure. Regarding course-level communication, more than half of the participants indicated that expectations regarding LLM use were communicated only in general terms. Specifically, 53.4\% (62 participants) reported receiving general guidance, whereas 23.3\% (27 participants) indicated that expectations were communicated clearly and explicitly. In contrast, 18.1\% (21 participants) reported that no guidance was provided, and 5.2\% (6 participants) did not recall whether expectations had been communicated. Participants were also asked whether potential consequences of inappropriate LLM use were discussed. A plurality of respondents 42.2\% (49 participants) reported that consequences were discussed briefly, while 25.9\% (30 participants) indicated that consequences were explained clearly. However, 20.7\% (24 participants) stated that consequences were not discussed at all, and 11.2\% (13 participants) were unsure. Finally, regarding the presence of institutional policies on AI or LLM use in coursework, 53.4\% (62 participants) reported that their institution had a formal policy. At the same time, 42.2\% (49 participants) indicated that they were not aware of any such policy, and 4.3\% (5 participants) were unsure. Collectively, these findings suggest substantial variability in how guidance, expectations, and policies related to LLM use are communicated to undergraduate students.

\subsection{Patterns of Perceived LLM Misuse Across Software Engineering Coursework}

We explored evidence on how software engineering students described using LLMs in ways they perceived as not aligned with course rules or expectations. Misuse was most frequently associated with \textbf{programming assignments or labs}, reported by 82 participants, corresponding to $\sim$71\% of the sample. Students commonly described using LLMs to generate code, debug errors, or complete substantial portions of assignments under time pressure or stress. These uses often exceeded the scope of conceptual assistance, even when course rules restricted LLMs to learning support. For example, one participant explained, \textit{``I have used it to help write some code since I was running out of time for the assignment and had other assignment to do and final exams coming up'' (P002)}, while another stated, \textit{``it was a python assignment and i only had 4 hours left so i had to use a LLM'' (P011)}. In some cases, participants explicitly acknowledged crossing clearly defined boundaries, as illustrated by P079: \textit{``I was stuck and worried about the deadline, so I asked the LLM to generate a full solution. I then edited the code and submitted it as my own work.''}

A substantial number of students also reported misuse in regular \textbf{classwork or weekly practices}, with $\sim$46\% (53 participants) of the sample, indicating this form of use. These accounts typically framed LLM use as a way to complete routine coursework more quickly, often with limited reflection on policy constraints. One participant noted, \textit{``I used an LLM to solve 100\% of my homework and quiz problems once'' (P006)}, while another stated, \textit{``I usually use an LLM to help me with my homework but this time I just copied the answers completely'' (P106)}.

Misuse in \textbf{software design or documentation tasks} was reported by 48 participants, corresponding to $\sim$41\% of the sample. Students described using LLMs to write documentation, generate reports, or produce design-related artifacts, often justifying this use by pointing to extensive requirements and limited time. For instance, one participant stated, \textit{``I used an LLM during an project to help write documentation… even though it was against the policy'' (P005)}, while another explained, \textit{``I made the documentation with CLaude's LLM… because we were running out of time and the requirements were huge'' (P098)}.

Use of LLMs for \textbf{essays or written reflections} was reported by $\sim$34\% of the sample (40 participants). In these cases, students often emphasized difficulties with organizing their ideas, insecurity about the quality of their writing, or a lack of interest in the assignment. One participant explained, \textit{``I use LLM for tasks where I struggle to organize my ideas… such as in reports'' (P021)}, while another stated, \textit{``I used an LLM for my assignment where I had to write an essay on digital leadership'' (P085)}. These responses suggest that LLMs were sometimes used as substitutes for individual writing rather than as tools for revision or feedback.

Misuse was also evident in \textbf{group or capstone projects}, corresponding to $\sim$33\% of the sample (38 participants). Students framed these decisions as pragmatic responses to collective deadlines, uneven workload distribution, or the risk of project failure. For example, one participant described copying group code into an LLM to locate bugs close to a deadline: \textit{``In the end I copied and pasted our work into ChatGPT and asked it to find the bugs. We were desperate at this point as the project was due in that week''. (P033)}. Another noted, \textit{``I used an LLM to help me write the mojority of the code… because i was the only one doing work so i neeeded help to finish in time'' (P113)}. These narratives indicate that collaborative contexts introduced additional pressures shaping LLM use.

A smaller proportion of participants reported misuse during \textbf{quizzes or short online tests}, with $\sim$29\% of the sample (34 participants), referencing this context. These accounts emphasized time constraints and grade pressure. One participant stated, \textit{``I used an LLM to complete a calculus quiz, thankfully my TA didn't notice'' (P024)}, while another noted, \textit{``Using LLM for a closed book quiz. The quiz is difficult and I needed the marks'' (P063)}. Although less frequent than assignment related misuse, these cases reflect direct boundary crossing in evaluative settings.

Finally, misuse during \textbf{major exams or finals} was reported by 17 participants, corresponding to $\sim$15\% of the sample. While comparatively less common, these accounts were often accompanied by explicit recognition of rule violations. For instance, one participant stated, \textit{``I used it during an exam, in which I felt I didn't have enough capacity'' (P017)}, while another explained, \textit{``I once used ChatGPT during an online exam… This went beyond what was allowed'' (P022)}.

\subsection{Conditions Surrounding Perceived LLM Misuse}

We investigated the factors that participants reported as enabling the use of LLMs in ways they perceived as misaligned with course rules or expectations. Across responses, pressures related to time and workload were most prominent. \textbf{Heavy workload or overlapping deadlines} was reported by 90 participants ($\sim$78\%) of the sample. In this context, students related LLM use as a pragmatic response to cumulative demands, competing deadlines, and sustained stress rather than as a deliberate attempt to violate rules. For example, one participant stated, \textit{``the pressure of deadlines was the biggest cause of me using the LLM on my assignments'' (P005)}, while another explained, \textit{``I was highly pressured and had multiple assignments due in a short time'' (P024)}.

In addition to workload pressures, the instructional context in which work was completed also shaped perceptions of misuse. \textbf{Assignments completed remotely or online} was selected by 70 participants ($\sim$60\%) of the sample. Participants described remote coursework and online assessments as lowering practical barriers to LLM use, particularly when deadlines were imminent. One participant explained, \textit{``I had fallen behind and didn't have time to completely start from scratch on a project that was due in 3 hours… going to school remotely helps the ease of use when needed'' (P013)}. Another explicitly linked online formats to reduced deterrence, stating, \textit{``With online assessment the fear of AI detection is much less present, as I can't be physically ‘caught’ ‘cheating’'' (P049)}.

Related to this context, some participants emphasized the role of monitoring and enforcement. \textbf{Limited oversight or checking by instructors} was reported by 43 participants ($\sim$37\%) of the sample. In these accounts, participants suggested that low visibility or weak checking practices reduced perceived risk and made LLM use feel easier to justify. One participant noted, \textit{``I needed to finish the work and the teacher isn't checking anything anyway'' (P008)}, while another reported, \textit{``For some maths quizzes, there was no way for them to tell'' (P033)}. A similar perception was reflected by P044, who stated, \textit{``I did not expect to be ‘caught’ because it wasn't even forbidden, so who knows if they even checked for LLM usage.''}

Uncertainty around institutional guidance further framed students' decisions. \textbf{Unclear or missing rules about AI use} was reported by 39 participants ($\sim$34\%) of the sample. Students described ambiguity regarding what constituted acceptable use, particularly for drafts, intermediate work, or coursework without explicit policy language. One participant explained, \textit{``Helping me to improve my learning of the material at hand was the primary motivation or unclear rules surrounding the use of LLMs for intermediate work'' (P003)}, while another reflected, \textit{``I believe the LLM use was technically a ‘grey’ area for my course'' (P019)}. Others pointed to the absence of formal guidance altogether, noting, \textit{``there wasn't any actual rules in the rubric or syllabus about LLM usage, unlike other courses'' (P067)}.

Alongside policy ambiguity, perceptions of risk played a distinct role. \textbf{Belief that AI use would not be detected} was also reported by 39 participants ($\sim$34\%) of the sample. These experiences emphasized a sense of low likelihood of consequences, particularly in online or low-stakes assessment contexts. For instance, one participant stated, \textit{``I just thought that I won't be detected'' (P075)}, while another noted that their use \textit{``went unnoticed by my learning institution'' (P049)}.

Finally, the social context contributed to decision-making for a substantial subset of participants. \textbf{Peer norms encouraging similar behavior} was selected by 35 participants ($\sim$30\%) of the sample. Participants described LLM use as normalized within their cohort, generating pressure to conform or remain competitive. One participant explained, \textit{``If everyone is using it, I will most probably do it as well'' (P018)}, while another stated, \textit{``I just think at least 80\% of my class is using it, so if I don't, then I am the dumb one'' (P038)}. Similarly, P095 noted, \textit{``Every single classmate is also using an LLM or an AI tool of some sort, so if I do not, I feel pressured to be left behind.''}

\subsection{Emotional Responses to Perceived Consequences of LLM Misuse}

We investigated how participants described their emotional responses when reflecting on the personal, academic, and professional consequences of relying on LLMs to bypass learning or assessment goals. Responses reveal five dominant affective orientations. These emotions range from concern and self reproach to detachment and, in a smaller number of cases, reassurance or perceived benefit.

\textbf{Anxiety about consequences} was reported by 21 participants ($\sim$18.1\%) of the sample. These responses were characterized by explicit concern about detection, sanctions, or long-term academic repercussions. Students referenced anxiety about grades, disciplinary actions, and institutional penalties. For example, one participant noted, \textit{``I think it could have a big impact on my grades if it is detected'' (P002)}, while another warned, \textit{``Failing the course, suspension and it going in my records'' (P076)}. More severe outcomes were also articulated, such as \textit{``At the worst being caught relying on LLM's can lead to permanent removal from the school and or department in which your major was'' (P013)} and \textit{``Chance of getting caught or flagged with the use of AI, resulting and losing the degree or having to redo courses'' (P095)}.

\textbf{Guilt} was reported by 24 participants ($\sim$20.7\%) of the sample. These responses reflected moral discomfort, self-blame, or a sense of having undermined their own education. Participants reporting guilt characterized LLM reliance as a personal failing rather than solely an institutional risk. One participant stated directly, \textit{``For personal, I think that guilt is a big consequence of relying on LLMs'' (P015)}. Others described internalized consequences such as diminished confidence, including \textit{``loss of credibility, lack of confidence in true skills and knowledge'' (P007)} and \textit{``The main consequence is a lack of self-confidence and the inability to truly learn fully'' (P021)}. Some participants questioned the meaning of their academic experience altogether: \textit{``What would be the point of me studying, receiving education at a university if I don't even learn anything'' (P084)}.

Following this, \textbf{Indifference} was the third frequently observed emotional stance, reported by 42 participants ($\sim$36.2\%) of the sample. These responses explicitly minimized or denied the existence of meaningful consequences. Several participants offered brief dismissals, such as \textit{``Mostly nothing happens'' (P018)}, \textit{``none to be honest'' (P072)}, and \textit{``None'' (P099 and P100)}. Others expressed certain levels of disengagement, stating \textit{``I don't know'' (P008)} or \textit{``I'm not sure'' (P062)}. A smaller subset rejected the premise of consequences altogether, asserting, \textit{``I think that there shouldnt be any consequences to begin with'' (P101)} or were framed as limited in scope, such as \textit{``I will have to retake the exam, nothing more'' (P034)}.

Subsequently, \textbf{Relief} was expressed by 21 participants ($\sim$18.1\%) of the sample. Their responses were marked by reassurance, rationalization, or perceived mitigation of risk. These students often qualified their views based on encouragement, context, or personal understanding. For instance, one participant stated, \textit{``I don't think there will be any consequence as I was encouraged to use LLMs'' (P053)}, while another emphasized conditional safety, noting, \textit{``Professionally, none, so long as I can understand what is being output and can verify it'' (P041)}. Others suggested that consequences depended on learning goals or circumstances, as in \textit{``I believe that it depends on the learning goal'' (P067)}.

Finally, \textbf{Satisfaction} was reported by 8 participants($\sim$6.9\%) of the sample. These responses reflected a positive attitude towards reliance on LLMs, even while acknowledging potential drawbacks. Participants emphasized efficiency, usefulness, or perceived improvement. One participant stated, \textit{``Using LLMs can significantly boost the learning ability of students'' (P047)}, while another described the impact broadly as \textit{``A huge improvement in tech'' (P092)}. Perceived gains in efficiency were also highlighted, as in \textit{``For me personally probably efficiency and speed depending on the topic'' (P030)}.

\subsection{Perceived Impacts of LLM Misuse on Software Engineering Education}

We investigated which aspects of software engineering education were perceived as most affected by the inappropriate use of LLMs. Concerns most frequently centered on \textbf{programming and implementation}, which was reported by 80 participants ($\sim$69\%) of the sample. These students associated LLM misuse with reduced hands-on practice and weakened implementation/coding skills. These responses demonstrate loss of experiential learning and diminished confidence in coding ability. One participant reflected, \textit{``Personally, it makes me a weaker programmer because I haven't spent enough time working hands-on with code'' (P039)}. Others described broader dependency concerns: \textit{``That you don't actually learn anything. Some of my peers have expressed worries about not actually knowing how to program without ai'' (P032)}. In more extreme terms, one participant summarized the outcome as, \textit{``Be an illiterate programmer'' (P063)}.

Closely related to programming, \textbf{software testing or debugging} was identified by 63 participants ($\sim$54.3\%) of the sample. These responses emphasized that misuse of LLMs could create difficulties in diagnosing errors and developing problem-solving strategies when such tools were used as substitutes rather than as supports. Participants described situations in which generated solutions undermined learning and conceptual understanding rather than facilitating them. One participant observed, \textit{``Many peers use LLMs for assignments, and then come to me for debugging. It is immediately obvious that they do not understand anything the LLM has produced'' (P031)}. Others stressed the loss of cognitive engagement inherent to debugging, stating, \textit{``They won't be used to actual thinking and trying to find a bug or problem on their on'' (P049)}. Uncertainty about validating outputs at scale was also raised: \textit{``when that scales how am I supposed to know when it's wrong?'' (P098)}.

Beyond code level activities, students also pointed out LLM misuse to impact on \textbf{technical documentation or reports}, which, as reported by 55 participants ($\sim$47.4\%) of the sample. These responses suggested that LLM misuse could reduce engagement with written technical artifacts and primary sources as they described risks associated with bypassing reading, synthesis, and documentation practices. One participant warned, \textit{``Not looking into source material and reading the papers, etc., could result in course failure and suspension'' (P111)}. Others associated documentation practices with reduced depth of engagement, noting, \textit{``There're examples where I've felt I haven't learned as much as I need for a topic in a business context from using LLMs to summarise or skip over large studies or whatever'' (P030)}.

Concerns about \textbf{individual skill assessments} were reported by 54 participants ($\sim$46.6\%) of the sample. These responses reflected perceived misalignment between assessment outcomes and actual competence. Students expressed discomfort when they experienced that the LLM misuse undermined their learning skills. One participant stated, \textit{``The main consequence is a lack of self-confidence and the inability to truly learn fully'' (P021)}. Another emphasized this disconnect more directly: \textit{``That I do not have the actual skills that are required for certain tasks'' (P044)}. This mismatch was also framed as cumulative across the degree, as one participant observed, \textit{``I feel I don't know how to do the tasks that i am supposed to know how to do as a third-year student'' (P115)}.

At a more conceptual level, \textbf{system design or architecture} was selected by 38 participants ($\sim$32.8\%) of the sample. These responses focused less on immediate outputs and more on understanding underlying structures and principles. Participants suggested that reliance on LLMs could undermine architectural reasoning and engagement with foundational concepts. One participant stated: \textit{``not understanding the architech and framework'' (P014)}. Another emphasized longer-term consequences: \textit{``It can lead to shallow understanding of core concepts, which can make later courses become significantly harder'' (P051)}.

Finally, \textbf{project management and teamwork} was identified by 37 participants ($\sim$31.9\%) of the sample. These responses emphasized coordination, time management, and effective contribution within collaborative settings. Participants associated LLM misuse with difficulties fulfilling responsibilities in group contexts and preparing for professional environments. One participant noted, \textit{``you may not be able to do the tasks assigned to you when you will be at work'' (P040)}, while another highlighted persistent skill gaps, stating, \textit{``I will still lack the skills in time management in my future'' (P046)}.

\subsection{Perceived and Anticipated Consequences of LLM Cheating}

Participants reported their understanding of the formal consequences that could follow if LLM cheating were identified, based on what they knew about their courses, departments, or institutions. The most frequently reported outcome was \textbf{course failure or zero grade}, cited by 44.8\% of participants (52 participants). A substantial proportion anticipated \textbf{formal warning or probation}, with 25.9\% (30 participants) reporting it. More severe institutional sanctions, such as \textbf{program suspension or expulsion}, were mentioned by 15.5\% (18 participants). In contrast, 13.8\% of participants (16 participants) believed that LLM cheating would result in \textbf{minor or no consequence}. These responses indicate that students associated LLM cheating with a range of academic penalties, most often situated at the assignment or course level rather than exclusively at the institutional level.

In addition to formal academic penalties, participants described expected consequences of LLM cheating that extended beyond direct course or assignment outcomes. The most commonly anticipated consequences were \textbf{reduced skill development} and \textbf{overreliance on AI tools}, each selected by 85.3\% of participants (99 participants). A smaller proportion anticipated affective consequences, including \textbf{feelings of dishonesty or discomfort}, reported by 34.5\% (40 participants), and \textbf{loss of trust from peers or instructors}, reported by 25.0\% (29 participants). A minority of respondents, 5.2\% (6 participants), indicated \textbf{no significant negative effect}. Altogether, these findings suggest that students see LLM cheating as having implications not only for formal assessment outcomes but also for learning trajectories, professional development, and academic relationships.

\section{Discussion} \label{sec:discussion}
In this section, we interpret the study findings in relation to the research question by synthesizing quantitative and qualitative results to identify recurring patterns across analyses. We then situate the observed patterns of LLM engagement in relation to prior studies on academic cheating and the use of LLMs in educational settings. Next, we discuss the implications of these findings for research and educational practice. Finally, we consider threats to validity that constrain the interpretation and transferability of the results.

\subsection{Interpreting LLM-Assisted Cheating in Software Engineering Education}

\begin{figure*}[!bth]
    \centering
    \includegraphics[width=0.9\textwidth]{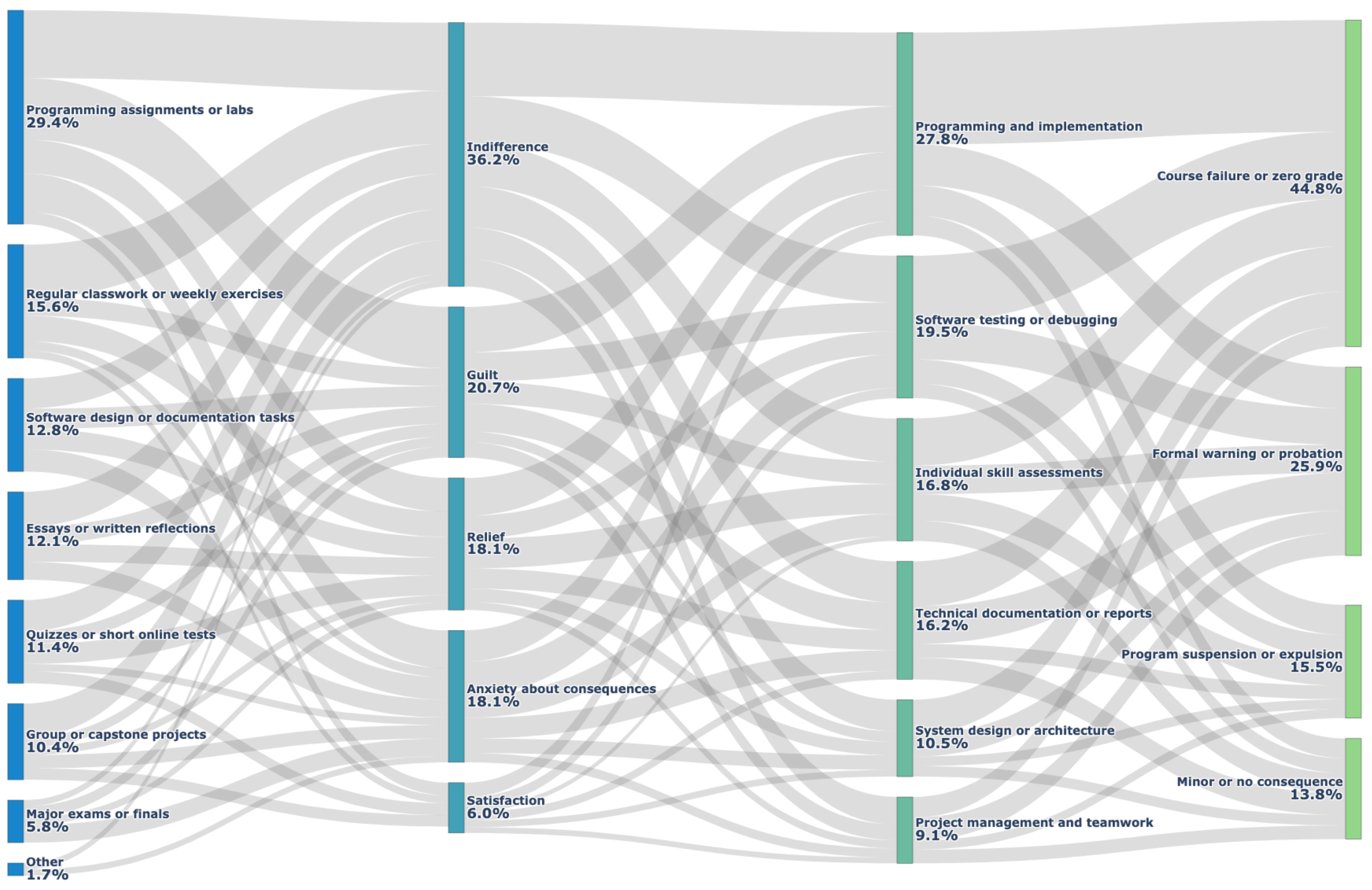}
    \caption{Weighted Sankey diagram illustrating the flow of student perceptions across stages of perceived LLM misuse, emotional responses, perceived impacts, and academic consequences. Node and link widths represent weighted proportions of respondents.}
    \label{fig:sankeyplot}
\end{figure*}

Figure \ref{fig:sankeyplot} supports a synthesized interpretation of how software engineering students relate the contexts of LLM use to their emotional responses, perceived learning effects, and anticipated consequences. Their accounts most often situate LLM misuse within routine coursework, particularly activities that recur across the term and demand sustained effort. Rather than presenting these moments as exceptional, they describe them as part of ordinary academic work, where pressures accumulate gradually. References to quizzes and major exams appear less central in their narratives, suggesting that they associate LLM misuse more closely with ongoing coursework than with clearly bounded evaluative events.

When reflecting on these practices, they most commonly describe an emotionally neutral stance, with indifference appearing more frequently than guilt, anxiety, or relief. This does not suggest that they perceive the behavior as inconsequential. Instead, emotional response appears loosely connected to how they assess the outcomes of LLM use. Across accounts, they converge on similar perceived impacts, particularly weakened engagement with programming, debugging, testing, and the development of individual competence. In this sense, how they feel about using LLMs varies, but how they judge its effects on learning remains largely aligned.

Moreover, students' reflections extend beyond learning impacts to expectations of academic consequences. Even when emotional responses are muted, they associate LLM misuse with substantive risks, including course-level penalties and institutional sanctions, alongside concerns about preparedness for professional practice. These anticipated consequences appear to be derived from their assessment of learning loss rather than from their emotional reactions, suggesting that awareness of risk is tied to judgments about educational outcomes rather than to how they feel about cheating using LLMs.

Finally, answering our research question (\textbf{RQ.} \textit{How do software engineering students describe their experiences with LLM assisted cheating and what contextual, motivational, and educational factors shape these behaviors?}), we can reflect that software engineering students describe LLM assisted cheating as a practice embedded in routine coursework and shaped by sustained workload, deadlines, and uncertainty around what would be the acceptable use. They present reliance on LLMs as a negotiated response to academic demands rather than as an impulsive or uninformed act. They differentiate assessment contexts, treating assignments and documentation as settings where LLM use is easier to rationalize, while recognizing clearer boundary violations in quizzes and exams. At the same time, they consistently acknowledge that these practices undermine learning and carry meaningful academic consequences, indicating that LLM cheating in software engineering education persists alongside an awareness of its educational and institutional implications.

\subsection{Comparing Findings}

The patterns observed in this study indicate that our findings align with long-standing research showing that cheating in higher education is rarely exceptional or purely individual, but instead emerges from contextual pressures, peer norms, and students’ interpretations of ambiguous rules~\cite{franklyn1995undergraduate, whitley1998factors, mccabe2001cheating}. Similar to earlier studies, students in our sample differentiated between practices they perceived as negotiable and those viewed as clearly prohibited, particularly across assessment types, reflecting variability in how formal rules are interpreted in practice~\cite{magnus2002tolerance, hosny2014attitude}. Our findings further show that workload intensity, overlapping deadlines, and low perceived likelihood of detection played a central role in shaping LLM-assisted cheating, consistent with prior work indicating that situational and social factors outweigh stable individual traits in explaining cheating behavior~\cite{whitley1998factors, bernardi2004examining, hutton2006understanding}. In addition, our observations are consistent with recent research on LLM use for cheating in other educational domains, which suggests that generative AI primarily reduces the effort required to produce acceptable outputs while increasing ambiguity around permissible assistance, rather than fundamentally altering students’ decision-making processes~\cite{cotton2024chatting, adnan2025cheating, lee2024cheating, ortiz2025chat}.

At the same time, our findings reveal features that distinguish software engineering from other academic areas. Software engineering education relies heavily on artifact-based assessment, where code, documentation, and integrated systems are treated as evidence of learning, despite reuse and automation being legitimate practices in professional contexts~\cite{sheard2002cheating, randall2023ai, marlowe2026systemic}. Students in our study most often described LLM-assisted cheating in programming assignments, debugging tasks, and documentation, particularly when policies specified acceptable outputs but did not clearly articulate expectations regarding development process or intermediate work. In this sense, our findings extend earlier research in software and computing education that documented persistent uncertainty around collaboration and reuse, by demonstrating how LLMs intensify these ambiguities through automated generation and modification of artifacts~\cite{sheard2002cheating, sheard2003investigating, salhofer2017analysing, rytilahti2024easy}. Conversely, in contrast to essay-based disciplines where LLM misuse is commonly framed in terms of authorship or originality, students in our study emphasized concerns related to bypassing core engineering activities, including programming, debugging, testing, and system reasoning, which are central to cumulative skill development in software engineering~\cite{salhofer2017analysing, rios2023authorship}.

In addition, our findings differentiate LLM-assisted cheating in software engineering from earlier forms of technology-mediated misconduct, such as plagiarism or unauthorized copying~\cite{jones2011academic, hsiao2015impact, spennemann2024chatgpt}. Prior work on digital cheating emphasized reuse of existing materials and similarity-based detection, whereas students in our study described generating novel artifacts that satisfied functional requirements without supporting learning~\cite{jones2011academic, hsiao2015impact, spennemann2024chatgpt}. Our findings also show that students were aware that such practices may undermine their educational trajectory and future professional competence, particularly in core activities such as programming and debugging, echoing concerns in the literature about diminished learning and misalignment between assessment outcomes and actual capability~\cite{whitley1998factors, cotton2024chatting, marlowe2026systemic}. These accounts suggest that LLM-assisted cheating in software engineering is sustained not by lack of awareness, but by tensions between performance-oriented coursework structures, ambiguous guidance, and expectations that students independently reconcile learning goals with increasing automation~\cite{randall2023ai, marlowe2026systemic}.

\subsection{Implications}

Our study offers implications for research and education that follow from software engineering students’ misuse of large language models.

\subsubsection{\textbf{Implications for Research}}

Our findings indicate a need for research approaches that move beyond measuring prevalence or tool capability and instead investigate how students interpret rules, negotiate ambiguity, and justify LLM use across different software engineering activities. The variation observed across programming, debugging, documentation, group projects, and exams suggests that future studies should adopt activity-sensitive designs rather than treating LLM cheating as a uniform phenomenon. In addition, the coexistence of awareness of long-term academic and professional consequences with continued misuse points to the need for explanatory models that account for the interaction between structural pressures, interpretive uncertainty, and short-term performance demands. Methodologically, our results support the use of mixed-methods designs that combine quantitative patterns with qualitative accounts of student reasoning, enabling more precise characterization of when and why boundary crossing occurs. Longitudinal research may further extend our work by investigating how these practices evolve as students progress through cumulative curricula and as institutional guidance stabilizes.

\subsubsection{\textbf{Implications for Education}}

Our findings suggest that educational responses to LLM cheating in software engineering should emphasize clarity, alignment, and attention to learning processes rather than relying primarily on prohibition or detection. Students’ accounts indicate that general or inconsistent guidance contributes to blurred interpretations of acceptable use, particularly for intermediate activities such as drafting or documentation. Instructional practice may therefore benefit from articulating expectations at the level of specific activities and processes, explicitly distinguishing learning-oriented support from substitution of core engineering work such as programming, testing, and architecture. In addition, the pragmatic framing of LLM misuse under heavy workload and time pressure suggests that assessment design and pacing meaningfully affect students' behavior. Approaches that incorporate iterative development, formative feedback, and process-based assessment may reduce incentives to bypass learning while remaining compatible with contemporary software engineering practice. Finally, students’ expressed concerns about diminished skill development and future preparedness highlight the importance of preserving opportunities for sustained hands-on engagement in programming, debugging, testing, and system reasoning across the curriculum.

\subsection{Threats to Validity} \label{sec:limitations}
As with other software engineering surveys, this study has limitations that should be considered when interpreting the findings~\cite{ralph2020empirical, pfleeger2001principles, linaker2015guidelines}. Participant recruitment relied on non-probability sampling and on self-reported information to identify software engineering students, which limits representativeness and introduces a risk of misclassification. To mitigate this limitation, recruitment on Prolific combined platform-level prescreening with survey-level screening questions that verified field of study, level of enrollment, and exposure to software engineering coursework and large language model use, and responses failing these checks were excluded during filtering. Snowball sampling was additionally used to extend participation beyond the platform. The study also relies on self-reported accounts of practices that participants themselves perceived as inappropriate or misaligned with course expectations, which may be affected by recall bias, social desirability, or variation in interpretation of academic integrity rules. These risks were addressed through anonymous data collection, careful question wording refined through piloting, and multiple data quality checks, including attention checks, completion time inspection, and manual review of open-ended responses for generic or automated content. Given this study design, our final sample does not support statistical generalization, as it draws on responses from 116 software engineering students, a sample size comparable to prior survey-based studies in software engineering involving both student and practitioner populations, but still insufficient for claims of population-level representativeness. The findings, therefore, rely on analytical generalization and transferability~\cite{ralph2020empirical}. Moreover, our study was not intended as a purely quantitative investigation, and qualitative data were collected to support interpretation and contextualization rather than inferential claims, as suggested in the guidelines~\cite{ralph2020empirical, linaker2015guidelines}.

\vspace{-0.5cm}
\section{Conclusions}
\label{sec:conclusion}

This study investigated software engineering students’ experiences with large language model use that they perceived as inappropriate or misaligned with course expectations, drawing on a cross sectional survey of 116 undergraduate participants. The findings indicate that LLM cheating is largely shaped by workload pressure, assessment structure, and unclear or inconsistent guidance, rather than by a lack of awareness of academic integrity norms. Such practices were most often reported in programming assignments, routine coursework, and documentation tasks, while use during quizzes and exams was less frequent and more clearly recognized as a violation. Students generally articulated awareness of potential academic and professional consequences, particularly reduced skill development and weakened confidence in core software engineering competencies. However, formal sanctions were often perceived as unlikely or limited, which appears to sustain boundary crossing under time pressure. These results suggest that responses focused on clearer alignment between learning goals, assessment practices, and explicit expectations for LLM use may be more effective than approaches centered solely on detection or prohibition. For future work, we plan to adopt longitudinal designs that follow students across core courses such as programming, testing, design, and capstone projects, to characterize how interpretations of acceptable LLM use evolve as curricular structures, assessment practices, and institutional guidance become more explicit and consistent.

\section*{Data Availability} \label{sec:DataAvailability}
The data analysed in this research is available at: \url{https://figshare.com/s/29c372b1260517aa0eb3}

\bibliographystyle{ACM-Reference-Format}
\bibliography{biblio}



\end{document}